\documentstyle[preprint,aps]{revtex}

\begin{document}

\title{Ab initio calculation of the thermal properties of Cu: Performance of the LDA and GGA}

\author{Shobhana Narasimhan}

\address{ Theoretical Sciences Unit, Jawaharlal Nehru Centre for
Advanced Scientific Research,\\ Jakkur PO, Bangalore 560 064, India}

\author{Stefano de Gironcoli}

\address{Scuola Internazionale Superiore di Studi Avanzati and
Istituto Nazionale per la Fisica della Materia,
via Beirut 2-4, I-34014 Trieste, Italy}

\maketitle

\begin{abstract} The thermal properties of bulk copper are investigated by
performing {\em ab initio} DFT and DFPT calculations and using the quasiharmonic
approximation for the free energy. Using both the LDA and the GGA for the
exchange-correlation potential, we compute the temperature dependence of the
lattice constant, coefficient of thermal expansion, bulk modulus,
pressure derivative of the bulk modulus, phonon frequencies,
Gr\"uneisen parameters, and the electronic and phonon contributions to the specific
heats at constant volume and constant pressure. We obtain answers in closer agreement
with experiment than those obtained from more approximate earlier treatments. The
LDA/GGA errors in computing anharmonic quantities are significantly smaller than those in 
harmonic quantities. We argue that this should be a general feature, and also
argue that LDA/GGA errors should increase with temperature.

\noindent {\it PACS \#}: 63.20.Ry, 65.40.+g, 65.70.+y, 71.15.Mb 

\end{abstract}

\newpage

\section{Introduction}
In any study of the properties of metals, it is obviously crucial to
include the effects of temperature. Thermal expansion results from the
anharmonicity of the interatomic potentials, and this change in the 
lattice constant upon heating a metal is accompanied by changes in the 
elastic and vibrational properties. Experimental measurements of
the temperature dependence of the lattice constant, elastic moduli, phonon
frequencies, Gr\"uneisen parameters, etc. of most elemental metals
have been available for a few
decades now. However, it has become possible to calculate these thermal
properties from first principles only in the last few years.

There are two main issues to be resolved when trying to compute the
thermal properties of metals: one is how to describe the interatomic
interactions accurately, and the other is how to incorporate the effects
of temperature into this description. 

To date, most computations of the thermal properties of metals have made
use of parametrized interatomic potentials.  
This necessarily introduces errors, even when the potentials are semi-empirical
and  include both theoretical and experimental values in the fitting database.
Self-consistent density functional theory (DFT) calculations
provide the most accurate way of computing interatomic interactions
from first principles. Using the DFT prescriptions to obtain the energies
as a function of nuclear co-ordinates avoids the errors introduced by
assuming parametrized forms of interatomic potentials.

As for computing the effects of temperature, one possible approach is to perform
molecular dynamics simulations at finite temperatures. This approach has,
for example, been combined with empirical and semi-empirical potentials to calculate
the thermal properties of metals. In principle, this approach can be extended
by performing {\em ab initio} molecular dynamics calculations at finite
temperatures for large unit cells containing many metal atoms.  
However, the amount of computational effort required in order to obtain
reliable thermodynamic averages makes this difficult, especially for the noble
metals and transition metals, which contain tightly-bound valence electrons. Also, since
such simulations treat the ionic degrees of freedom classically, the results
are not valid at very low temperatures, when zero-point effects are important.

An alternative approach is to compute the vibrational free energy using
the quasiharmonic approximation, in which anharmonic effects are included
via the volume-dependence of phonon frequencies, which can be determined by performing
{\em ab initio} calculations. Here too, in order to perform
reliable averages, it is necessary to compute the frequencies for many wave-vectors
in the Brillouin zone (BZ), which is computationally expensive, especially if the
phonon frequencies are calculated using the ``frozen-phonon" method. 
The development of density-functional perturbation theory (DFPT)
\cite{dfpt} has considerably 
reduced the computational cost of obtaining phonon frequencies throughout the
BZ, since unlike the frozen phonon method, this technique does not require using
large supercells to access wave-vectors away from the zone center.

Thus, combining {\em ab initio} DFPT calculations with a quasiharmonic
treatment of the anharmonicity of vibrations currently offers us the most
reliable yet practicable approach towards calculating averaged thermal
properties, at least up to temperatures not too close to the melting
point. In recent years, this combined approach has been shown to be quite
successful in predicting the bulk thermal properties of the simple metals
Al, Li and Na \cite{quong}, and the noble metal Ag \cite{xie}.

However, there remains one important issue that has to be decided 
when performing {\em ab initio} calculations:  how to describe the exchange and correlation
effects in the electron-electron interactions. The exact form of the exchange-correlation
functional is not known, and one has to use various approximate schemes; the most
widely used ones being the local density approximation (LDA) and various versions of   
generalized gradient approximations (GGAs). The GGAs are intended to be an 
improvement on the conventional LDA, and do indeed perform better in certain situations,
such as transition states in chemical reactions, or systems containing ``weak" bonds.
Unfortunately, however, the GGAs do not always give answers that are in better
agreement with experiment. 

Improving the treatment of exchange and correlation effects
is the holy grail in the field of electronic structure calculations,
and as an aid towards achieving this goal, it is desirable to have a clear picture of the 
comparative merits of the LDA and GGA in various situations. 
It has been known for a long time now that the LDA tends
to ``overbind", giving lattice constants that are too small, and bulk moduli, phonon
frequencies and cohesive energies that are too large. The GGAs seem to overcorrect these
errors,
giving lattice constants that are too large. A recent study \cite{favot} showed
that this overcorrection is  manifested also in the harmonic properties: 
the GGA gives bulk
moduli and phonon frequencies that are systematically lower than the experimental ones.
We are not aware of any detailed studies comparing the performance of the LDA and GGA
in describing anharmonic effects, which manifest themselves in   
the temperature-dependence of the lattice constant, elastic and
vibrational properties, and specific heat capacities, and in the values of anharmonic
quantities such as the Gr\"uneisen parameters.

To this end, in this paper, we have performed {\em ab initio} calculations to study
the thermal properties of bulk copper, using both the LDA and GGA.
We have computed the temperature-dependence
of the lattice constant, the coefficient of thermal expansion, the isothermal
bulk modulus, the phonon frequencies,
the individual and overall Gr\"uneisen parameters, and the specific heat capacities at constant volume and
constant pressure.

\section {Ab initio calculations}

The {\em ab initio} calculations were performed using the PWSCF 
and PHONON codes \cite{pwscf}.
Total energies were computed using DFT, and phonon frequencies using DPFT. The
interaction between the ions and valence electrons was described using an
ultrasoft pseudopotential \cite{ultrasoft}. A plane-wave basis set with a cut-off of 30 Ry was used;
a cut-off of 300 Ry was used in the expansion of the augmentation charges necessitated
by the use of the ultrasoft (non norm-conserving) pseudodopotential. Brillouin-zone
integrations were performed using 60 {\bf k} points in the irreducible part of the
BZ. Phonon dynamical matrices were computed {\em ab initio} for a 
$4 \times 4 \times 4$ {\bf q} point mesh; Fourier interpolation was then 
used to obtain the dynamical matrices on a $24 \times 24 \times 24$ {\bf q} point mesh. 
This latter set was used to evaluate
all quantities that involve an integration over phonon wave-vectors {\bf q}.

In order to deal with the possible convergence problems for metals, a smearing
technique was employed using the Methfessel-Paxton (MP) scheme \cite{methfessel}, 
with the smearing
parameter $\sigma$ set equal to 0.05 Ry. However, when evaluating the electronic
contribution to the specific heat capacity (as explained below) we instead used
a Fermi-Dirac (FD) smearing, with the electronic levels occupied according to the
FD distribution appropriate to the temperature of interest $T$. (Incidentally, with
this latter scheme, we did not face convergence problems even at low values of $T$.)

When using the LDA, we used the parametrization by Perdew and Zunger of the results
of Ceperley and Alder \cite{pz}. For the GGA,  we used the Perdew-Burke-Ernzerhof form
\cite{pbe};
this choice was made in part because it is easier to implement in the DPFT calculations,
and because it gives a good description of the linear response of the uniform
electron gas.

To summarize, the following were obtained from DFT and DPFT calculations:
(i) Total energies at a range of lattice constants, using (a) MP smearing
(b) FD smearing, for a range of temperatures between 1 K and 1400 K.
(ii) For each lattice constant, the dynamical matrices (and thus phonon frequencies) 
for the $4 \times 4 \times 4$ set of {\bf q} points, using MP smearing; Fourier
interpolation was then used to obtain the dynamical matrices on the 
$24 \times 24 \times 24$ set of {\bf q} points. 
(It was verified that replacing the MP
smearing by the FD smearing did not make an appreciable difference to the phonon
frequencies, i.e., the latter are not sensitive to the electronic temperature.) 
All of the above quantities were computed using both the LDA and the GGA.
This set of results was then used to calculate the thermal behavior, as described below. 

\section{Results and Analysis}

The static results for lattice constant $a_0$, the bulk modulus $B_0$ and the
pressure-derivative of the bulk modulus $B^\prime$ are obtained by fitting
the results for the static total energies (using MP smearing) versus lattice constant to
the fourth order Birch-Murnaghan equation of state \cite{birch}. 
Using the LDA, we obtain  $a_0$ = 6.71 bohr,
$B_0$ = 1.72 MBar, and $B^\prime$ = 5.0. The corresponding results with the
GGA are $a_0$ = 6.94 bohr, $B_0$ = 1.28 MBar, and $B^\prime$ = 5.11. As
expected, the experimental values for the lattice constant ($a_0$ = 6.82 bohr) 
\cite{ashcroft}
and bulk modulus (1.37 Mbar) \cite{kittel} lie sandwiched between the LDA and GGA values;
it should however be noted that the experimental values are at room temperature,
and the calculated values listed above do not yet include the effects of temperature.
For $B^\prime$, there does not seem to be a consensus on the experimental value,
with several values reported in the literature. Listed in chronological
order, these are: 3.91\cite{lazarus}, 5.3\cite{bridgman}, 4.8\cite{altshuler},
4.1\cite{rice}, 5.59\cite{daniels} and 5.44\cite{hiki}.

To study the effects of changing temperature, one has to look at the free energy,
incorporating the effects of thermal vibrations (phonons).
The free energy at temperature $T$ and lattice constant $a$ is given, within
the quasiharmonic approximation, by:

\begin{equation}\label{eq:freeen} F(a,T)  = E_{\rm stat}(a) +
k_B T \sum_{{\bf q}\lambda} ln \Biggl\{2 {\rm sinh} \Biggl({{\hbar\omega_{{\bf
q}\lambda}(a)}\over{2 k_B T}}\Biggr)\Biggr\}  . \end{equation} 

\noindent
Here, the first term on the right-hand-side is the static energy $E_{\rm stat}(a)$,
and the second term is the vibrational free energy. The sum is over all three phonon
branches $\lambda$ and over all wave-vectors {\bf q} in the BZ (we will use
the $24 \times 24 \times 24$ {\bf q} mesh in evaluating this),
 $\hbar$ is Planck's constant, $k_B$ is Boltzmann's
constant, and 
$\omega_{{\bf q}\lambda}(a)$ is the frequency of the phonon with wave-vector
{\bf q} and polarization $\lambda$, evaluated at lattice constant $a$.

The lattice constant at temperature $T$, $a_0(T)$ is obtained by minimizing
$F(a,T)$ with respect to $a$. The linear expansion $\epsilon(T)$ is then given
by 

\begin{equation}\label{eq:epsilon} \epsilon(T) = {{a_0(T)-a_0(T_c)}
\over{a_0(T_c)}},
\end{equation}

\noindent
where $T_c$ is the reference temperature of 298.15 K.

Fig.~1 shows the results for $\epsilon(T)$,
(expressed as a percentage) using both the LDA and the GGA, compared to the experimental 
value \cite{aip}. It is seen
that the agreement with experiment is quite good, though the LDA slightly
underestimates the expansion, and the GGA slightly overestimates it.
This becomes more obvious upon differentiating the results for $a_0(T)$ to obtain
the coefficient of linear expansion: 

\begin{equation}\label{eq:alpha}
\alpha(T)={1 \over a_0(T_c)} {\Biggl({da_0(T)\over dT}\Biggr)}.
\end{equation}

\noindent
(Note that this definition of $\alpha(T)$ is the one used for experimental data.
When using $\alpha(T)$ in thermodynamic relations, $a_0(T_c)$ should be replaced
by $a_0(T)$ in the right-hand-side of the above equation.)
Fig.~2 compares the calculated and experimental values \cite{aip} for $\alpha(T)$ 
up to a temperature of 1400 K (the experimental value for the bulk melting temperature
is 1357 K). Once again, it is clear that the experimental values lie sandwiched
between the LDA and GGA values, though they lie somewhat closer to the LDA values,
especially at high temperature. However, it should be pointed out that the calculated 
values may be inaccurate at very high temperatures for two reasons: (i) The use
of the quasiharmonic approximation may not be justified at temperatures 
just below the melting point, as
this is expected to be a region of high anharmonicity. (ii) A part of the experimentally
measured thermal expansion at high temperatures results from the formation of
vacancies; this effect is not included in our calculations, where we assume that
the crystal remains defect-free at all temperatures.

By fitting the results for the free energy from Equation 1
to the fourth-order Birch-Murnaghan equation of state \cite{birch}, 
we also obtain the variation in temperature
of the bulk modulus $B_0$ and the pressure derivative of the
bulk modulus $B^\prime$. We find that the quality of the fit is
noticeably better with the fourth-order equation of state than with the
Murnaghan equation of state\cite{murnaghan} or with the
third-order Birch-Murnaghan equation, especially at higher temperatures.
The results for $B_0(T)$ are  plotted in Fig.~3,
from which it can be seen that though, at all temperatures, the absolute value of 
$B_0(T)$ is overestimated by the LDA and underestimated by the GGA, the rate of
change of $B_0$ with temperature is approximately the same for both, and moreover,
this rate agrees well with that measured experimentally \cite{chang}.
Fig.~4 shows the results for $B^\prime(T)$; it can be seen that $B^\prime$ depends
noticeably on the temperature. (Incidentally this temperature dependence
is considerably underestimated if one uses the Murnaghan equation or the
third-order Birch-Murnaghan equation.) At 300 K, the LDA and GGA values for
$B^\prime$ are 5.21 and 5.40, respectively, compared to the static values of
5.00 and 5.11. These values agree well with some of the
room-temperature experimental values cited
above \cite{lazarus,bridgman,altshuler,rice,daniels,hiki}, 
though there is a considerable scatter in the experimentally reported
values.

Since we know how the phonon frequencies vary with $a_0$, and how $a_0$ 
varies with $T$, it is now a simple matter to get the phonon frequencies at
any desired temperature. In Fig.~5, the calculated and measured 
\cite{cuphon} phonon frequencies,
at a temperature $T$=80 K, are plotted along several high-symmetry directions in
the Brillouin zone. At this temperature, the LDA gives $a_0$ = 6.73 bohr,
and the GGA gives $a_0$ = 6.96 bohr (since the temperature is relatively low,
there is not an appreciable change from the static values). Yet again, it can be
seen that the experimental values lie in between the LDA and GGA values. The
over/under-estimation of the frequencies by the LDA/GGA can be traced back to
the under/over-estimation of the lattice constant. In fact, if the phonon frequencies
are computed at the {\em experimental} lattice constant, the situation is reversed,
and the GGA frequencies are {\em higher} and the LDA frequencies are {\em lower} than
experiment, though the latter are closer to the experimental values than the former.

We can also compute the temperature-dependence of the specific heat capacities
at constant volume and constant pressure, as described below. The specific heat
at constant volume has two contributions, one from the phonons, and the other from
the electrons. The former is given by

\begin{equation}\label{eq:cvph}
C_V^{\rm ph}(T) = \sum_{{\bf q}\lambda}C_v({\bf q}\lambda) = k_B \sum_{{\bf q}\lambda} 
\Biggl({{\hbar\omega_{{\bf q}\lambda}(a_0(T))}\over{2 k_B T}}\Biggr)^2
{\rm sinh}^{-2} \Biggl({{\hbar\omega_{{\bf q}\lambda}(a_0(T))}\over{2 k_B T}}\Biggr).
\end{equation}

The electronic contribution to the specific heat, $C_V^{\rm el}(T)$ is obtained from
the self-consistent DFT calculations using FD smearing corresponding to a temperature
$T$, by computing the derivative with respect to the smearing temperature $T$ of the
electronic entropy, evaluated at the corresponding lattice constant $a_0(T)$.
The total specific heat at constant volume is then $C_V^{\rm tot}(T) =
C_V^{\rm ph}(T) + C_V^{\rm el}(T)$. 

$C_p$, the specific heat at constant pressure, can then be computed by using the
relation:

\begin{equation}\label{eq:cp}
C_p(T) = C_V^{\rm tot}(T) + {9\over 4} \alpha^2(T)B_0(T)a_0(T) T.
\end{equation}

Figs.~6 (a) and (b) show the results thus obtained for $C_V^{\rm ph}(T)$, $C_V^{\rm el}(T)$,
$C_V^{\rm tot}(T)$ and $C_p(T)$, computed using the LDA and GGA respectively. 
As expected, the electronic contribution to the specific heat, $C_V^{\rm el}(T)$,
is much smaller than the phonon contribution $C_V^{\rm ph}(T)$, though not
negligible. 
The experimental values for $C_p(T)$ \cite{aip} are also plotted. It is seen that for
both the LDA and GGA, the agreement with experiment is excellent up to about
600 K. Above this temperature, the agreement remains very good for the LDA,
but is poorer for the GGA. Note that at these high temperatures, 
$C_V^{\rm ph}(T)$ has reached its saturation value of 3$k_B$ per atom, and the LDA and
GGA values for $C_V^{\rm ph}(T)$ are therefore identical. For the difference
between $C_p$ and $C_V$, the error due to the under/over-estimation of $\alpha$ 
by the LDA/GGA is to
some extent cancelled out by the over/under-estimation of $B_0$. 

The anharmonicity of the vibrations can be examined by computing the mode 
Gr\"uneisen parameters, defined by

\begin{equation}\label{eq:gruen} 
\gamma_{{\bf q}\lambda} = - {V \over \omega_{{\bf q}\lambda}(V)}
{\partial \omega_{{\bf q}\lambda}(V) \over \partial V}
,
\end{equation}

\noindent
where $V = a^3/4$ is the volume of the unit cell. Fig.~7 shows the results for
the Gr\"uneisen parameters for the same high-symmetry modes for which the frequencies
were plotted in Fig.~5. They have been evaluated at the static lattice constants.
Though the LDA and GGA static lattice constants are different, it can be seen
that the discrepancy in the corresponding Gr\"uneisen parameters is small; considerably
smaller than the differences in phonon frequencies. For
example, at the X point (zone edge
along [100]), the discrepancy between the LDA and GGA results for the phonon
frequencies is 10.7\% and 12.5\% for the transverse and longitudinal branches
respectively, whereas the corresponding Gr\"unesien parameters differ by
only 0.4\%
and 2.1\% respectively.
It is interesting to compare Fig.~7 with Fig.~3 of Ref.~\onlinecite{xie}, which shows the
corresponding result for Ag. Though the phonon dispersion curves of Ag and Cu are
very similar in shape and structure, our curves for the 
Gr\"uneisen parameters of Cu look quite different, in some areas of the BZ,
from those reported earlier for Ag.

We also compute 
the overall Gr\"uneisen parameter $\gamma$, which is obtained by
averaging over the individual Gr\"uneisen parameters 
$\gamma_{{\bf q}\lambda}$ of all the modes, using the equation:

\begin{equation}\label{eq:gam_avg}
\gamma(T) = {{\sum_{{\bf q}\lambda} \gamma_{{\bf q}\lambda} C_V({\bf q}\lambda)}
            \over
            {\sum_{{\bf q}\lambda} C_V({\bf q}\lambda)} }  .
\end{equation}

\noindent
where the contribution from each mode $({\bf q}\lambda)$ is weighted by $C_V({\bf q}\lambda)$,
its contribution to the specific heat, as defined in Equation 4. This quantity is
of interest because it appears in some useful thermodynamic relations (as discussed
below) and experimental papers often report this overall value. The temperature
dependence of $\gamma$ comes from the temperature dependence of both the individual
Gr\"uneisen parameters $\gamma_{{\bf q}\lambda}$ (which depend on the lattice constant
and hence on T) and that of the specific heat. 

Fig.~8 shows the results for the results for the variation of the overall
Gr\"uneisen parameter $\gamma$ as a function of temperature. It is seen that the
percentage difference between the LDA and GGA results is small, ranging from
2.5\% at 100 K to 5.3\% at 1300 K. The discrepancy with experiment is also
quite small, with the LDA and GGA errors being 2.8\% and 4.8\% respectively
at room temperature. In comparison, the LDA and GGA results for the bulk modulus
(plotted in Fig.~3) differ from each other by 30 to 50\% over the same temperature
range, with the LDA and GGA errors (with respect to experiment) being 18.8\%
and 13.7\% respectively at room temperature. 

\section{Comparison with earlier calculations}

We compare our results to those of three previous calculations. In the
first, interatomic potentials are described by a pair potential fit to experimental
data, and thermal effects are treated
by formulae that are valid in the high-temperature limit. In the second,
the static energies at zero temperature are computed {\em ab initio}; however,
thermal effects are computed using a Debye model and various approximate relations;
i.e., the phonon frequencies are not calculated {\em ab initio}.
In the third calculation, the interatomic potentials are described by an 
empirical form that includes some many body effects, and thermal effects are treated 
using finite-temperature molecular dynamics simulations. 

In their study of the thermodynamic properties of face-centered-cubic (FCC) metals,
MacDonald and MacDonald \cite{macdonalds}
have described the interatomic interactions using a modified
Morse potential fit to experimental data such as the Debye temperature $\Theta_D$, the sublimation
energy, and the thermal expansion in the neighborhood of $\Theta_D$
(342 K).
The electronic contribution to $C_V$ is estimated using free-electron theory.
Though the thermal expansion is fit to agree with experiment
at low temperatures, $\alpha$ is underestimated
by about 20\% at 1200 K (in comparison, our results for $\alpha$ at 1200 K, with no fit to
data on thermal expansion, differ from experiment by -12\% when using the LDA and +29\% when
using the GGA.) The absolute value of $B_0$ agrees well with experiment
(which is to be expected, given the fitting to $\Theta_D$); however 
$\partial B_0(T)/\partial T$ is underestimated. The magnitude of the electronic contribution
to $C_V$, estimated from free-electron theory, is similar to what we obtain from our
more exact approach; and the agreement between the calculated and experimental values for
$C_p$ is fairly good, similar to that obtained by us. 
The value of $\gamma$ changes from 1.947 at 100 K to 2.127 at 1000 K, which is comparable
to our results. However, it should of course be
kept in mind that in our calculations we do not fit to any empirical data at all.

Moruzzi, Janak and Schwarz \cite{moruzzi} have computed the bulk binding curve
(i.e., $E_{\rm stat}(a)$) by
performing {\em ab initio} augmented-spherical-wave method calculations. They use this to obtain
$B_0$ (and thus an approximate $\Theta_D$) and $\gamma$ (independent of $T$). 
The free energy is then
evaluated in the Debye model, with the volume-dependence of the frequencies being
determined by $\gamma$. They evaluate $\alpha$ only up to $T = 300$ K, getting a
value of $\alpha$ that is too low at 300 K by 20\%. Our calculation improves
upon this one in that we do not use a Debye model, and do not assume that all modes
have the same degree of anharmonicity, since we calculate individually and exactly
the values of $\omega_{{\bf q}\lambda}(a)$. This is probably
why our calculated values for $\alpha$ are closer to experiment: at 300 K, our errors
in the calculated $\alpha$ are -14\% (LDA) and +14\% (GGA).

\c Ca\u gin {\em et al.} \cite{cagin} have used the empirical Sutton-Chen potential to describe the interatomic
interactions. The potential parameters are fit to the cohesive energy, bulk modulus, etc.,
at 0 K. Temperature effects are determined by performing molecular dynamics simulations.
As a comparison, let us consider the values for the thermal expansion $\epsilon$ at 1000 K:
they obtain a result of 2.42\%, compared to our values of 1.22\% (LDA) and 1.64\% (GGA),
and the experimental value of 1.37\%. Our values are clearly closer to experiment; in this
case, their larger errors presumably arise from deficiencies in their interatomic
potential.

To summarize: though our results do not agree exactly with experiment, we still do a 
better job than earlier calculations. This is because we have eliminated
the errors due to the utilization of parametrized interatomic potentials 
and an approximate treatment of the
lattice vibrations.  The (smaller) errors that remain in our calculations are due
to the choice of exchange-correlation potential (and, possibly, the use of the quasiharmonic
approximation, though we believe these errors to be small).

\section{Discussion of results}

From the results presented in Section III, it is clear that the 
approximations used for the 
exchange-correlation potentials  introduce much smaller errors in 
anharmonic quantities such as $\gamma$, $B^\prime$ and $\partial B_0(T)/\partial T$, 
than in harmonic properties like the
bulk modulus and phonon frequencies. 

At first sight, the relatively large
discrepancy between the LDA and GGA values for the coefficient of thermal expansion
$\alpha$ may seem to contradict this statement.
However, $\alpha$ is not a purely anharmonic quantity.  For example,
in a one-dimensional anharmonic potential given by $V(x) = {1\over 2}cx^2 - 
{1\over 6} gx^3$, 
we have $\alpha \propto g/c^2$ \cite{kittel}, i.e., $\alpha$ depends on both the harmonic coefficient
$c$ and the anharmonic coefficient $g$, and an error in the former will be manifested
as an even larger error in $\alpha$. For the present case, where we have to average
over a number of normal modes in three dimensions, it is possible to derive \cite{ashcroft} a 
corresponding equation relating $\alpha$ to the averaged anharmonic quantity
$\gamma$ and harmonic quantity $B_0$:

\begin{equation}\label{eq:alpha_form}
\alpha(T) = {\gamma(T) C_V(T) \over 3B_0(T)},
\end{equation}

From Fig.~8, it is seen that the discrepancy between the LDA and GGA values of
$\gamma(T)$ is small, and that both are close to experiment. Any error in
$C_V(T)$ is negligible, especially at temperatures above the Debye temperature,
where $C_V$ has reached its saturation value of $3k_B$ per atom. There remains
the large error in $B_0(T)$. At a temperature of 1000 K, for example, the
LDA and GGA values for $\gamma$, $B_0$ and $\alpha$ differ by 4\%, 41\%, and 32\%
respectively, which is consistent with our argument that the discrepancy
in $\alpha$ arises almost entirely from the discrepancy in $B_0$.

If we assume that the main source of the LDA and GGA errors in the values of
physical quantities is the wrong value obtained for the lattice constant, then it
is indeed consistent that the errors in anharmonic quantities should be smaller
than those in harmonic quantities. For example, let us return to the example
of the one-dimensional cubic anharmonic potential cited above:

\begin{equation}\label{eq:anharmpot}
V(a) = {1\over 2}c(a-a_0)^2 - {1\over 6} g(a-a_0)^3  .
\end{equation}

\noindent
If the second derivative (which gives harmonic properties) is evaluated 
at a lattice constant $a$ which is not the
correct lattice constant $a_0$, the error in the second derivative  
is given by ${\partial^2V / \partial a^2}\mid_a - c = -g(a-a_0)$, and the
greater the error in the lattice constant, the greater the error in the
calculated harmonic properties. However,
there is no error in the third derivative (which gives cubic anharmonic
properties), since ${\partial^3V / \partial a^3}\mid_a = -g$ for all $a$.
(If one were to add a quartic term to Eq.~9, then the error in the second
derivative would have terms linear and quadratic in $(a-a_0)$, whereas the error in the
third derivative would be proportional to $(a-a_0)$, and the fourth
derivative would be exact.) Thus, errors arising from using a wrong lattice
constant are manifested to lesser and lesser degrees as one goes to higher
order anharmonic properties.

One can also argue that the LDA and GGA errors made in computing physical properties should
increase with temperature: At T=0, the LDA underestimates $a_0$ and
the GGA overestimates it. Upon heating, the LDA underestimates the thermal expansion
(because of the overestimation of $B_0$) and the GGA overestimates it (because of
the underestimation of $B_0$). Thus, as the temperature is increased, the underestimation by the
LDA and the overestimation by the GGA of the lattice constant are both aggravated
further,
resulting in increasingly unreliable results for physical properties, though the
errors are smaller for anharmonic properties than harmonic ones. To avoid this,
we suggest that when performing {\em ab initio} DFT calculations at high temperatures
in systems that display large errors in the calculated bulk modulus,
it is perhaps a good idea to use the experimental values of the thermal expansion,
regardless of whether one is using the LDA or GGA. It is also useful to keep in mind
that the error in the static value for the bulk modulus is already a good indicator
of the magnitude of the error that will be made in the coefficient of thermal expansion;
thus, if in a particular case, 
either the LDA or the GGA gives a better value for ths static
value of $a_0$ and $B_0$, it will probably
also give a better description of finite-temperature properties.

\section{Summary}
To summarize: we have performed {\em ab initio} calculations to study the thermal properties
of bulk copper, using phonon frequencies computed using DFPT, and the quasiharmonic
approximation for the  vibrational free energy. We have calculated the temperature
dependence of the lattice constant $a_0$, the thermal expansion
$\epsilon$, the coefficient of thermal expansion $\alpha$, the bulk modulus $B_0$,
the pressure derivative of the bulk modulus $B^\prime$,
the overall Gr\"uneisen parameter $\gamma$, and the lattice and electronic
contributions to the specific heat capacities and constant volume and pressure,
$C_V$ and $C_p$. We have also presented results for the dispersion, along high-symmetry
directions in the BZ, of the phonon frequencies and mode Gr\"uneisen parameters. All of
the above have been computed using both the local density approximation (LDA) and generalized
gradient approximation (GGA).

Neither the LDA nor the GGA is clearly to be preferred in this case, with both giving
errors of comparable magnitude (though generally of opposite sign).
At all temperatures, the LDA systematically underestimates the lattice constant and the
coefficient of thermal expansion, and the GGA overestimates these. In contrast, the LDA
always overestimates the bulk modulus and phonon frequencies, and the GGA 
underestimates them. The electronic contribution to the specific heat is found to be
considerably smaller than the phononic contribution, though not neglible.
The results for $C_p$ agree very well with experiment, except for
the GGA results at high temperatures. However, the discrepancy between the
LDA and GGA results (and their discrepancy with experiment) is considerably lower for
anharmonic quantities such as $\gamma$, $B^\prime$ and $\partial B_0(T)/\partial T$
than for the harmonic properties. In any event, our results are closer to experiment
than those of earlier calculations in which the interatomic interactions and/or
thermal effects were treated approximately.

We have argued that if the main source of errors can be attributed to the wrong value
obtained for the lattice constant resulting from the approximate nature of the
exchange-correlation potential, then it is indeed reasonable that the errors in
anharmonic quantities should be smaller than those in harmonic quantities. We have also
argued that the LDA/GGA errors should increase with temperature, suggesting the need
for care and caution when performing {\em ab initio} calculations at high temperatures.

\section{Acknowledgments}

SN gratefully acknowledges support from the Associateship program of
the Abdus-Salam International Centre for Theoretical Physics, Trieste,
Italy, that made this collaboration possible.
SdG has been supported  in part by MURST under COFIN99, and by the {\sl
Iniziativa
Trasversale Calcolo Parallelo}  of INFM.


\begin{figure}
\caption{Linear thermal expansion $\epsilon$ as a function of temperature, referred to a
reference temperature $T_c$ of 298.15 K. The experimental
values are from Ref.~\protect\onlinecite{aip}. It is seen that both the LDA (solid line) and
GGA (dashed line) results are close to the experimental values.}
\end{figure}

\begin{figure}
\caption{Coefficient of linear thermal expansion $\alpha$ as a function of temperature.
Both the LDA and GGA values are reasonably close to the experimental values; however the
LDA underestimates and the GGA overestimates the thermal expansion. Experimental values
are from Ref.~\protect\onlinecite{aip}.}
\end{figure}

\begin{figure}
\caption{Variation with temperature of the bulk modulus $B_0$. At all temperatures,
the LDA (solid line) overestimates $B_0$ and the GGA (dashed line) underestimates it;
however, $\partial B_0/\partial T$ is approximately the same for the LDA, the GGA,
and the experimental values. The experimental values are from Ref. \protect\onlinecite{chang}.}
\end{figure}

\begin{figure}
\caption{Variation with temperature of the pressure derivative of the bulk
modulus, B$^\prime$. The solid line is the LDA result, and the dashed line is
the GGA result.} 
\end{figure}

\begin{figure}
\caption{Phonon dispersion along high-symmetry directions in the BZ, at 80 K. The solid
and dashed lines are the results obtained using the LDA and GGA respectively, and the
filled circles are the experimental values from Ref. \protect\onlinecite{cuphon}.
``L" and ``T"
denote the longitudinal and transverse branches respectively.}
\end{figure}
 
\begin{figure}
\caption{Calculated values of $C_p$ and $C_v$, in units of $k_B$ per atom, obtained
using the (a) LDA and (b) GGA. The dot-dashed lines show $C_V^{\rm ph}$, the thin
dashed lines show $C_V^{\rm el}$, and the thick dashed lines show their sum
$C_V^{\rm tot}$. The solid lines show the calculated values for $C_p$, obtained from
$C_V^{\rm tot}$ by using Eq.~(5). The dots show the experimental results for
$C_p$, as given in Ref.\protect\onlinecite{aip}.}
\end{figure}

\begin{figure}
\caption{ Calculated  dispersion curves
for the individual mode Gr\"uneisen parameters $\gamma_{{\bf q}\lambda}$ for the
same high-symmetry directions in the BZ as in Fig.~5. The values have been evaluated
at the static lattice constant; the discrepancy between the LDA (solid lines) and
GGA (dashed lines) results is small. ``L" and ``T"
denote the longitudinal and transverse branches respectively.}
\end{figure}

\begin{figure}
\caption{ Overall Gr\"uneisen parameter $\gamma$ as a function of temperature.
The solid and dashed lines are the results obtained using the LDA and GGA respectively.
The experimental value is taken from Ref.\protect\onlinecite{moruzzi}. }
\end{figure}

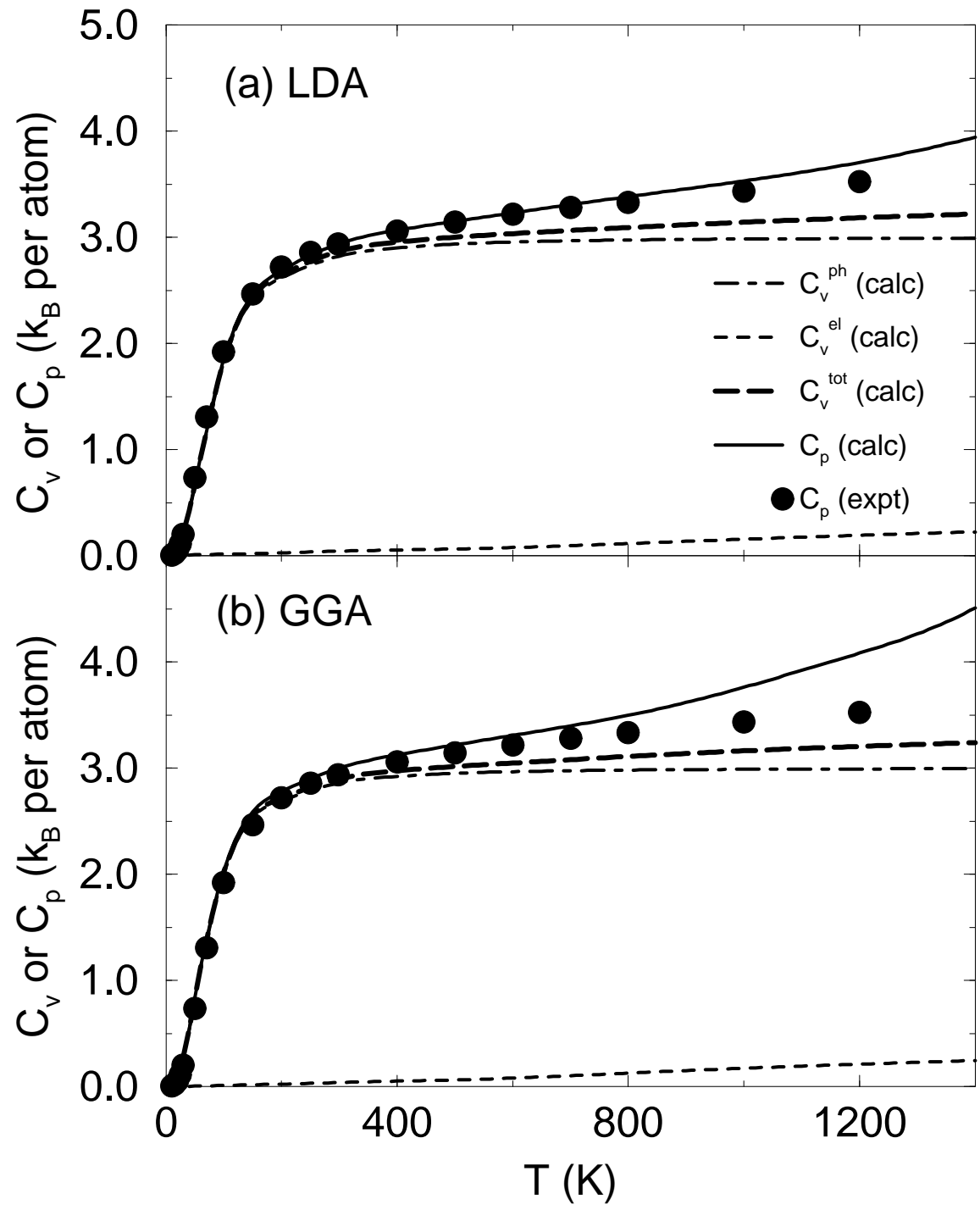
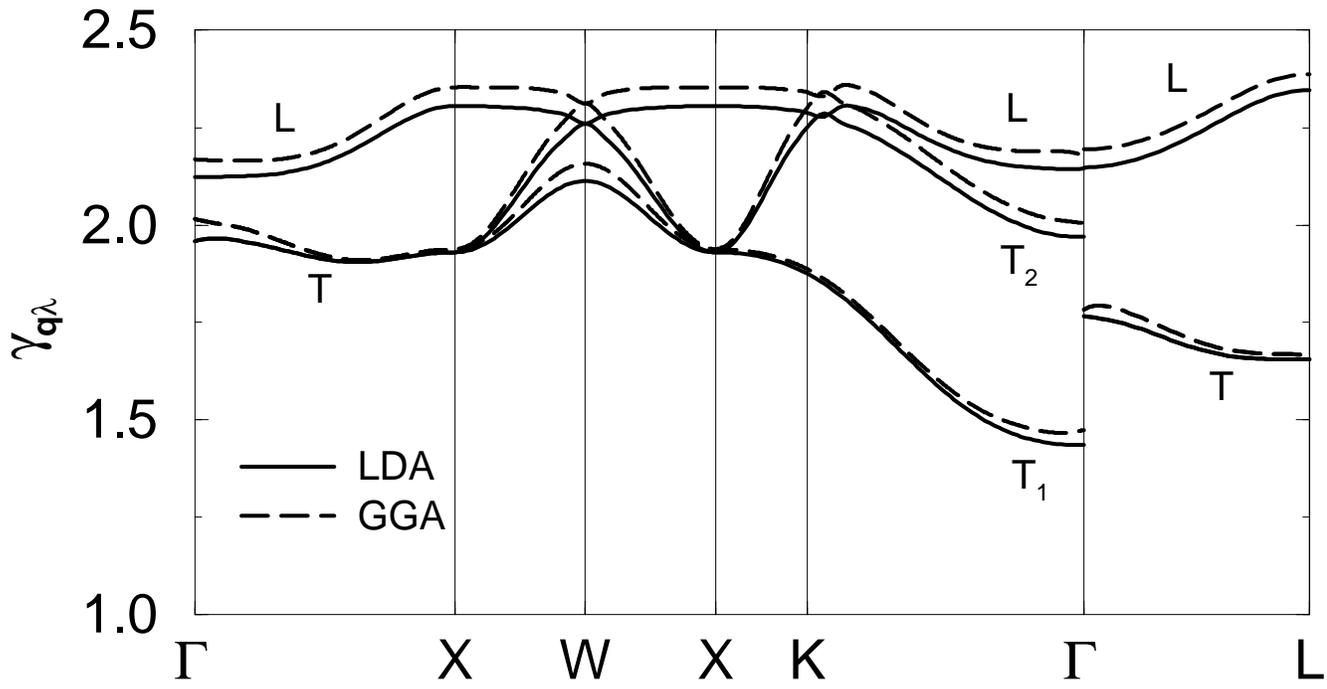
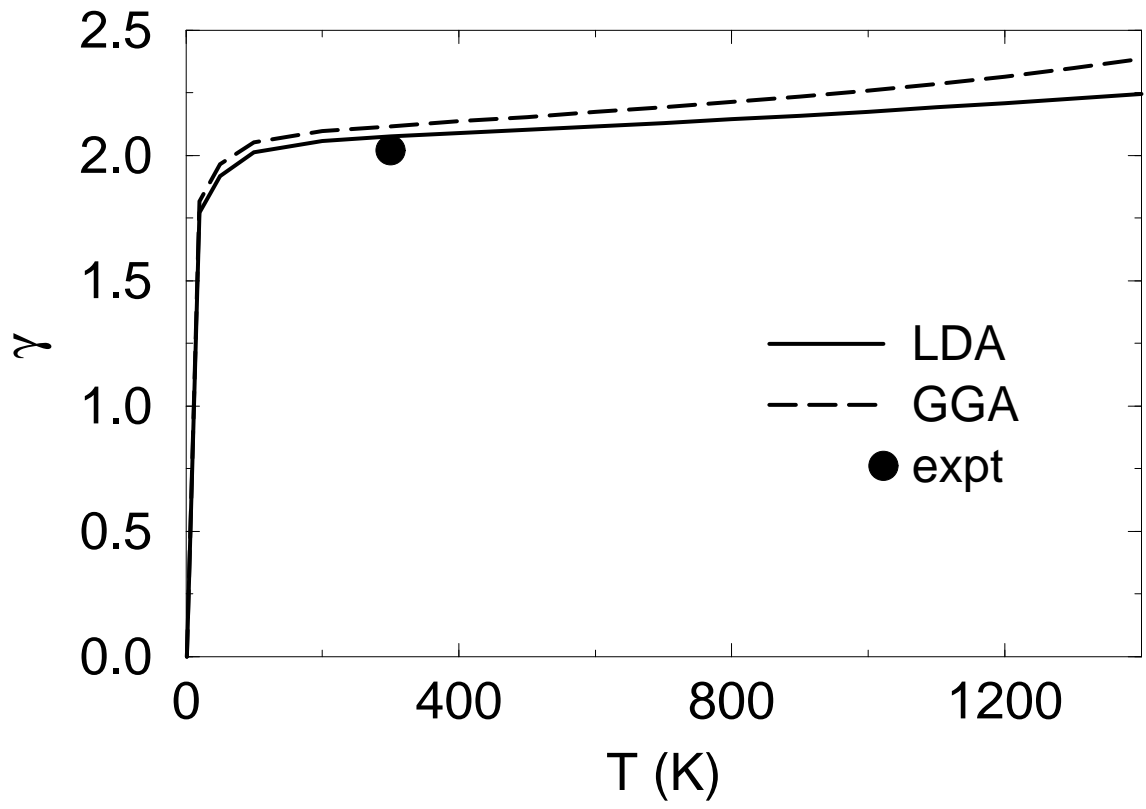
\end{document}